# Algebraic theory of linear viscoelastic nematodynamics

# Part 1: Algebra of nematic operators

Arkady I. Leonov

*Department of Polymer Engineering, The University of Akron,*

*Akron, Ohio 44325-0301, USA.*

**Abstract**

This *first part* of the paper develops algebraic theory of linear anisotropic, six-parametric nematic "N-operators" build up on the additive group of traceless second rank 3D tensors. These operators have been implicitly used in continual theories of nematic liquid crystals and weakly elastic nematic elastomers. It is shown that there exists a non-commutative, multiplicative group $\mathbb{N}_6$ of N-operators build up on a manifold in 6D space of parameters. Positive N-operators, which in physical applications holds thermodynamic stability constraints, form a subgroup of group $\mathbb{N}_6$ on a more complicated manifold in parametric space. A three-parametric, commutative transversal-isotropic subgroup $\mathbb{S}_3 \subset \mathbb{N}_6$ of positive symmetric nematic operators is also briefly discussed. The special case of singular, non-negative symmetric N-operators reveals the algebraic structure of *nematic soft deformation modes*.

## Introduction

One of the central problems in theoretical descriptions of liquid crystalline polymers (LCP) is a lack of general continual (or ”field”) theory, which could consistently describe their specific viscoelastic nematic properties. It seems that this situation is caused by using an awkward common tensor/matrix form of operations in nematic theories (e.g. see [1-3]), which does not allow displaying their simple algebraic structure. It makes difficult (if possible) elaborating a general theory of linear or weakly nonlinear dynamic behavior of LCP’s.

This paper reveals the algebraic structure of the continual nematic theories and presents it in a simple form. This will allow us to demonstrate in the second part of the paper some remarkable features of linear nematic viscoealsticity of LCP.

### 1.1. Definitions and general properties

Consider a set $\breve{X}$ of traceless 3D second rank Cartesian tensors $\underline{\underline{x}} = \{x_{ij}\} \in \breve{X}: \ tr\underline{\underline{x}} = 0.$ The set $\breve{X}$ is forms an additive group defined on the field of real numbers. The group $\breve{X}$ is naturally decomposed in the sum, $\breve{X} = \breve{X}_s + \breve{X}_a$, $(\breve{X}_s \bigcap \breve{X}_a = 0)$ of two additive subgroups of symmetric $\breve{X}_s$ and asymmetric $\breve{X}_a$ matrices, so that $\forall \underline{\underline{x}} \in \breve{X}: \ \underline{\underline{x}} = \underline{\underline{x}}_s + \underline{\underline{x}}_a$, where $\underline{\underline{x}}_s \in \breve{X}_s$ and $\underline{\underline{x}}_a \in \breve{X}_a$.

We introduce on $\breve{X}$ linear, axially symmetric operations transforming $\breve{X}$ into itself. The axial symmetry is characterized by a given unit vector $\underline{n}$ (*director*), disposed arbitrarily relative to axes of a chosen Cartesian coordinate system. A linear operation, invariant relative to transformation $\underline{n} \to -\underline{n}$, is called *nematic operation* (or simply N-operator). The implicit definition of N-operation in the common tensor presentation is:

$$\underline{\underline{y}}_s = r_0 \underline{\underline{x}}_s + r_1[\underline{\underline{nn}} \cdot \underline{\underline{x}}_s + \underline{\underline{x}}_s \cdot \underline{\underline{nn}} - 2nn(\underline{\underline{x}}_s : \underline{\underline{nn}})] + r_2(\underline{\underline{nn}} - \underline{\underline{\delta}}/3)(\underline{\underline{x}}_s : \underline{\underline{nn}}) + r_3(\underline{\underline{nn}} \cdot \underline{\underline{x}}_a - \underline{\underline{x}}_a \cdot \underline{\underline{nn}})$$

$$\underline{\underline{y}}_a = r_4(\underline{\underline{nn}} \cdot \underline{\underline{x}}_s - \underline{\underline{x}}_s \cdot \underline{\underline{nn}}) - r_5(\underline{\underline{nn}} \cdot \underline{\underline{x}}_a + \underline{\underline{x}}_a \cdot \underline{\underline{nn}}). \qquad (1.1)$$

Here $(\underline{\underline{nn}})_{ij} = n_i n_j$, the symbol “$\cdot$” means the tensor (matrix) multiplication, “:” the trace operation, $\underline{\underline{\delta}}$ is the unit tensor, and $r_k$ are the six independent *ordered basis parameters*, characterizing the operation in (1.1), denoted as: $\mathrm{r} = (r_0, r_1, ..., r_5)$. It is seen that $\underset{=s}{y} \in \breve{X}_s, \underset{=a}{y} \in \breve{X}_a$, i.e. the operation (1.1) transforms $\breve{X} \to \breve{X}$. Relations (1.1) have been used as constitutive equations for viscous and weakly elastic nematic cases [1-5]. The possible term ~ $\underset{=}{x}^a$ was excluded from (1.1) due to physical arguments (e.g. see [3-5]),

We now denote the N-operator as $\mathbf{N}_r(\underline{n})$, and symbolically present (1.1) as $\forall \underset{=}{x}, \underset{=}{y} \in \breve{X}: \ \underset{=}{y} = \mathbf{N}_r(\underline{n}) \bullet \underset{=}{x}$. Although the parameters $r_k$ are generally independent, there is still one physically significant, "Onsager" case of their relation, $r_4 = -r_3$. So we define the *Onsager* N-*operator* (or ON-operator) as: $\mathbf{N}_r^o(\underline{n}) \equiv \mathbf{N}_r(\underline{n})\Big|_{r_4=-r_3}$.

For any N-operator $\mathbf{N}_r(\underline{n})$ we can introduce on $\breve{X}$ the quadratic form, a scalar $P$, defined as:

$$P \equiv \underset{=}{x} \bullet \mathbf{N}_r(\underline{n}) \bullet \underset{=}{x} \equiv r_0 \left|\underset{=}{x}^s\right|^2 + 2r_1 \underline{n}\underline{n} : \underset{=s}{x}^2 + (r_2 - 2r_1)(\underline{n}\underline{n} : \underset{=s}{x})^2 - 4r_3^* \underline{n}\underline{n} : (\underset{=s}{x} \cdot \underset{=a}{x}) - 2r_5 \underline{n}\underline{n} : \underset{=a}{x}^2 . \quad (1.2)$$

$$r_3^* \equiv (r_3 - r_4)/2 .$$

In case of ON operator $\mathbf{N}_r^o(\underline{n})$, when $P \to P^o$, the quadratic form $P^o$ has potential properties: $2\underset{=s}{y} = \partial P^o / \partial \underset{=s}{x}$, $2\underset{=a}{y} = \partial P^o / \partial \underset{=a}{x}$.

Operator $\mathbf{N}_r(\underline{n})$ is called *positive* if $\forall \underset{=}{x} \in \breve{X}: \ P \equiv \underset{=}{x} \bullet \mathbf{N}_r(\underline{n}) \bullet \underset{=}{x} > 0$. The same holds for ON operator $\mathbf{N}_r^o(\underline{n})$.

*Theorem 1.*

(i) The N-operator $\mathbf{N}_r(\underline{n})$ is positive ($P > 0$) iif:

$$\mathrm{r} \in \tilde{R}_6^+: \ r_0 > 0; \quad r_0 + r_1 > 0; \quad 3/2 r_0 + r_2 > 0; \ (r_0 + r_1) r_5 > (r_3 - r_4)^2 / 4 \equiv r_3^{*2} . \quad (1.3_{1,2,3,4})$$

(ii) Any positive N-operator $\mathbf{N}_r(\underline{n})$ has inverse, $\mathbf{N}_r^{-1}(\underline{n})$.

*Proof*

Using an orthogonal transformation, we choose the coordinate system, whose axis 1 is directed along the director. In this coordinate system (1.1,2), written in the component form, are reduced to:

$$y_{11}^s = (r_0 + 2/3 r_2) x_{11}; \ y_{22}^s = r_0 x_{22} - x_{11} r_2 / 3; \ y_{33}^s = r_0 x_{33} - x_{11} r_2 / 3; \ y_{23}^s = r_0 x_{23} \quad (1.1_{1a})$$

$$\begin{cases} y_{1k}^s = (r_0 + r_1) x_{1k}^s + r_3 x_{1k}^a \\ y_{1k}^a = -r_4 x_{1k}^s + r_5 x_{1k}^a \end{cases} \quad (k = 2,3) \quad (1.1_{1b})$$

$$P=(3/2r_0+r_2)x_{11}^2+2r_0[(x_{22}+x_{11}/2)^2+x_{s23}^2]+2\sum_{k=2,3}[(r_0+r_1)x_{s1k}^2+2r_3^*x_{a1k}x_{s1k}+r_5x_{a1k}^2]\quad(1.2_1)$$

Here the traceless condition, $x_{11}+x_{22}+x_{33}=0$, has been used to exclude $x_{33}$ from (1.2$_1$). Demanding $P>0$ and using independence of terms in (1.2) yields inequalities (1.3). Note that the inequality (1.3$_4$) yields the inequality:

$$(r_0+r_1)r_5>-r_3r_4\,.\qquad(1.3._5)$$

When (1.3$_{1,2,3,5}$) holds for equations (1.1$_1$), there is a unique linear dependence of $x_{ij}$ on $y_{ij}$, which means the existence of a unique inverse operation $\mathbf{N}_r^{-1}(\underline{n})$.

*Remark 1.1*

Theorem 1 holds for ON-operators $\mathbf{N}_r^o(\underline{n})$ when $r_3^*=r_3\ \ (P\to P^o)$.

To make inverse resolution of equations (1.1) it is not necessary to use in the parametric space, $\mathrm{r}\in R_6$, the manifold $\tilde{R}_6^+$ with inequalities (1.3$_{1\text{-}4}$). The necessary and sufficient conditions for this resolution are: $r_0\neq 0,\ \ 3/2r_0+r_2\neq 0,\ (r_0+r_1)r_5+r_3r_4\neq 0$. Nevertheless, it is more convenient to use in the following the positive conditions of the resolution:

$$\mathrm{r}\in R_6^+:\ \ r_0>0;\quad 3/2r_0+r_2>0;\ \ (r_0+r_1)r_5+r_3r_4>0\,.\qquad(1.3_{1,3,5})$$

N-operator is called *$N^+$-operator* if its basis parameters satisfy less restrictive inequalities (1.3$_{1,3,5}$) than (1.3$_{1\text{-}4}$), valid for positive N-operators. Evidently, $\tilde{R}_6^+\subseteq R_6^+$, i.e. $N^+$-operators are not necessarily positive though any positive N-operator is $N^+$-operator.

## 1.2. Basis N-operators and their multiplicative properties

The tensor/matrix presentations [6] of *basis* N-operators $\mathbf{a}_k(\underline{n})$ (k=0,1,...,5) in (1) is explicitly defined via fourth rank numerical tensors (or simply 4-tensors) $\{\mathbf{a}_k(n)\}_{ij\alpha\beta}$ as:

$$\{\mathbf{a}_0\}_{ij\alpha\beta}=a_{ij\alpha\beta}^{(0)}=1/2(\delta_{i\alpha}\delta_{j\beta}+\delta_{i\beta}\delta_{j\alpha}-2/3\delta_{ij}\delta_{\alpha\beta})\qquad(1.4_1)$$

$$\{\mathbf{a}_1(\underline{n})\}_{ij\alpha\beta}=a_{ij\alpha\beta}^{(1)}=1/2(\delta_{i\alpha}^{\perp}n_jn_\beta+\delta_{j\alpha}^{\perp}n_in_\beta+\delta_{i\beta}^{\perp}n_jn_\alpha+\delta_{j\beta}^{\perp}n_in_\alpha)\quad \delta_{ij}^{\perp}=\delta_{ij}-n_in_j\qquad(1.4_2)$$

$$\{\mathbf{a}_2(\underline{n})\}_{ij\alpha\beta}=a_{ij\alpha\beta}^{(2)}=(n_in_j-1/3\delta_{ij})(n_\alpha n_\beta-1/3\delta_{\alpha\beta})\qquad(14_3)$$

$$\{\mathbf{a}_3(\underline{n})\}_{ij\alpha\beta}=a_{ij\alpha\beta}^{(3)}=1/2(\delta_{i\alpha}n_jn_\beta+\delta_{j\alpha}n_in_\beta-\delta_{i\beta}n_jn_\alpha-\delta_{j\beta}n_in_\alpha)\qquad(1.4_4)$$

$$\{\mathbf{a}_4(\underline{n})\}_{ij\alpha\beta} = a^{(4)}_{ij\alpha\beta}(\underline{n}) = 1/2(\delta_{i\alpha}n_j n_\beta + \delta_{i\beta}n_j n_\alpha - \delta_{j\alpha}n_i n_\beta - \delta_{j\beta}n_i n_\alpha) = \{\mathbf{a}_3(\underline{n})\}_{\alpha\beta ij} \qquad (1.4_5)$$

$$\{\mathbf{a}_5(\underline{n})\}_{ij\alpha\beta} = a^{(5)}_{ij\alpha\beta}(\underline{n}) = 1/2(\delta_{i\beta}n_j n_\alpha - \delta_{i\alpha}n_j n_\beta + \delta_{j\alpha}n_i n_\beta - \delta_{j\beta}n_i n_\alpha) \qquad (1.4_6)$$

The basis 4-tensors in ($4_{1-6}$), are traceless with respect of the first and the second pairs of indices, i.e. $\{\mathbf{a}_k(\underline{n})\}_{ii\alpha\beta} = \{\mathbf{a}_k(\underline{n})\}_{ij\alpha\alpha} = 0$ $(k = 0,..,5)$.

The following symmetry properties hold for the basis 4-tensors:

$$\{\mathbf{a}_k(\underline{n})\}_{ij\alpha\beta} = \{\mathbf{a}_k(\underline{n})\}_{ji\alpha\beta} = \{\mathbf{a}_k(\underline{n})\}_{ij\beta\alpha} = \{\mathbf{a}_k(\underline{n})\}_{\alpha\beta ij} \qquad (k = 0,1,2) \qquad (1.5_1)$$

$$\{\mathbf{a}_3(\underline{n})\}_{ij\alpha\beta} = \{\mathbf{a}_3(\underline{n})\}_{ji\alpha\beta} = -\{\mathbf{a}_3(\underline{n})\}_{ij\beta\alpha} \qquad (1.5_2)$$

$$\{\mathbf{a}_4(\underline{n})\}_{ij\alpha\beta} = -\{\mathbf{a}_3(\underline{n})\}_{ji\alpha\beta} = \{\mathbf{a}_3(\underline{n})\}_{ij\beta\alpha} \qquad (1.5_3)$$

$$\{\mathbf{a}_5(\underline{n})\}_{ij\alpha\beta} = -\{\mathbf{a}_5(\underline{n})\}_{ji\alpha\beta} = -\{\mathbf{a}_5(\underline{n})\}_{ij\beta\alpha} = \{\mathbf{a}_5(\underline{n})\}_{\alpha\beta ij}\,. \qquad (1.5_4)$$

Formulae ($1.4_{1-3}$) and ($1.5_1$) show that the 4-tensors $\{\mathbf{a}_k(\underline{n})\}_{ij\alpha\beta}$ (k = 0,1,2) are symmetric relative to transposition of the first and second indices, the third and fourth indices, as well as the first and second pairs of indices; the tensor $\{\mathbf{a}_3(\underline{n})\}_{ij\alpha\beta}$ is symmetric relative to transposition of the first and second indices, and skew symmetric when transposing the third and fourth indices; the tensor $\{\mathbf{a}_4(\underline{n})\}_{ij\alpha\beta}$ is asymmetric relative to transposition of the first and second indices, and symmetric when transposing the third and fourth indices; and the tensor $\{\mathbf{a}_5(\underline{n})\}_{ij\alpha\beta}$ is skew symmetric relative to transposition of the first and second, as well as of the third and fourth indices, and symmetric when transposing the first and second pairs of indices.

The symmetry properties of the above 4-tensors $\mathbf{a}_k(\underline{n})$ $(k = 0,..,5)$ show that they represent irreducible (and therefore linearly independent) set of traceless 4-th rank tensors. Therefore thy are called *basis* tensors.

The products of the tensors of different ranks are defined as:

$$\mathbf{a}_s(\underline{n}) \bullet \mathbf{a}_r(\underline{n}) \Rightarrow a^{(s)}_{ij\alpha\beta} a^{(r)}_{\beta\alpha\nu\gamma}, \quad \mathbf{a}_r(\underline{n}) \bullet \underline{\underline{x}} \Rightarrow a^{(r)}_{ij\alpha\beta} x_{\beta\alpha}. \qquad (1.6)$$

The definitions in (6) disclose the sense of operation "•" symbolically used in Section 2. The products $\mathbf{a}_s(\underline{n}) \bullet \mathbf{a}_r(\underline{n})$, established directly are presented in Table 1.

*Table 1. Products of basis tensors* $\mathbf{a}_i \bullet \mathbf{a}_j$

| $i$ \ $j$ | 0 | 1 | 2 | 3 | 4 | 5 |
|---|---|---|---|---|---|---|
| 0 | $\mathbf{a}_0$ | $\mathbf{a}_1$ | $\mathbf{a}_2$ | $\mathbf{a}_3$ | $\mathbf{0}$ | $\mathbf{0}$ |
| 1 | $\mathbf{a}_1$ | $\mathbf{a}_1$ | $\mathbf{0}$ | $\mathbf{a}_3$ | $\mathbf{0}$ | $\mathbf{0}$ |
| 2 | $\mathbf{a}_2$ | $\mathbf{0}$ | $(2/3)\,\mathbf{a}_2$ | $\mathbf{0}$ | $\mathbf{0}$ | $\mathbf{0}$ |
| 3 | $\mathbf{0}$ | $\mathbf{0}$ | $\mathbf{0}$ | $\mathbf{0}$ | $-\mathbf{a}_1$ | $\mathbf{a}_3$ |
| 4 | $\mathbf{a}_4$ | $\mathbf{a}_4$ | $\mathbf{0}$ | $-\mathbf{a}_5$ | $\mathbf{0}$ | $\mathbf{0}$ |
| 5 | $\mathbf{0}$ | $\mathbf{0}$ | $\mathbf{0}$ | $\mathbf{0}$ | $\mathbf{a}_4$ | $\mathbf{a}_5$ |

It is seen that except $i,j$ = 0,1,2, the multiplication of basis tensors is non-commutative, for example, $\mathbf{a}_0 \bullet \mathbf{a}_3 = \mathbf{a}_3 \neq \mathbf{a}_3 \bullet \mathbf{a}_0 = \mathbf{0}$ .

### 1.3. Multiplicative group of N-operators

Using basis operators $\mathbf{a}_k(\underline{n})$, equation (1.1) can be rewritten in the operator form:

$$\underline{\underline{y}} = \mathbf{N}_r(\underline{n}) \bullet \underline{\underline{x}}, \qquad \mathbf{N}_r(\underline{n}) \equiv \sum_{k=0}^{5} r_k \mathbf{a}_k(\underline{n}), \tag{1.7$_1$}$$

or equivalently as:

$$\underline{\underline{y}}_s = \sum_{k=0}^{2} r_k \mathbf{a}_k(\underline{n}) \bullet \underline{\underline{x}}_s + r_3 \mathbf{a}_3(\underline{n}) \bullet \underline{\underline{x}}_a, \qquad \underline{\underline{y}}_a = r_4 \mathbf{a}_4(\underline{n}) \bullet \underline{\underline{x}}_s + r_5 \mathbf{a}_5(\underline{n}) \bullet \underline{\underline{x}}_a \ . \tag{1.7$_2$}$$

The *product* of two N-operators is defined in the common way:

$$\mathbf{N}_p(\underline{n}) \equiv \mathbf{N}_q(\underline{n}) \bullet \mathbf{N}_r(\underline{n}) = \sum_{k,m=0}^{5} q_k r_m \mathbf{a}_k(\underline{n}) \bullet \mathbf{a}_m(\underline{n}) = \sum_{k=0}^{6} p_k \mathbf{a}_k(\underline{n}). \tag{1.8}$$

With the use of multiplicative Table 1, the basic scalars $p_k$ for resulting operation are found from the *fundamental* equation:

$$\begin{aligned} &p_0 = q_0 r_0, \ \ p_1 = q_0 r_1 + q_1 r_0 + q_1 r_1 - q_3 r_4, \ \ p_2 = q_2 r_0 + q_0 r_2 + 2/3 q_2 r_2 \\ &p_3 = (q_0 + q_1) r_3 + q_3 r_5, \ \ p_4 = q_4 (r_0 + r_1) + q_5 r_4, \ \ p_5 = -q_4 r_3 + q_5 r_5 \end{aligned} \tag{1.9}$$

Even in the Onsager case, when $q_4 = -q_3$ and $r_4 = -r_3$, generally $p_4 \neq -p_3$. It means that $\mathbf{N}_p(\underline{n}) \neq \mathbf{N}_p^o(\underline{n})$ i.e. that the product of two ON-operators is not an ON-operator.

*Theorem 2*

The set of $N^+$-operators $\mathbf{N}_r(\underline{n})$, whose basis parameters $\mathtt{r} \in R_6^+$ satisfy inequalities ($1.3_{1,3,5}$), constitute a non-commutative, multiplicative, six-parametric group $\mathbb{N}_6$, which has the fundamental group equations (1.9).

*Proof*

(i) The definition and basic properties of the *unit* N-operator $\mathbf{I}(\underline{n})$ due to the Table 1 are:

$$\mathbf{I}(\underline{n}) = \mathbf{a}_0 + \mathbf{a}_5(\underline{n}), \quad \mathbf{N}_r(\underline{n})\bullet\mathbf{I}(\underline{n}) = \mathbf{I}(\underline{n})\bullet\mathbf{N}_r(\underline{n}) = \mathbf{N}_r(\underline{n}) \tag{1.10}$$

Because of (1.2),($1.2_1$) the unit N-operator is positive, and therefore it is a $N^+$-operator.

(ii) If $\mathbf{N}_r(\underline{n})$ is $N^+$-operator, its inverse, $\mathbf{N}_r^{-1}(\underline{n}) \equiv \mathbf{N}_{\hat{r}}(\underline{n})$, existing due to theorem 1, should satisfy the common condition, $\mathbf{N}_{\hat{r}}(\underline{n})\bullet\mathbf{N}_r(\underline{n}) = \mathbf{N}_r(\underline{n})\bullet\mathbf{N}_{\hat{r}}(\underline{n}) = \mathbf{I}(\underline{n})$ which for parameters in (9) yields:

$$p_0 = 1, \;\; p_1 = p_2 = p_3 = p_4 = 0, \;\; p_5 = 1. \tag{1.11}$$

The basis parameters $\hat{r}_k$ of inverse N-operator $\mathbf{N}_{\hat{r}}(\underline{n})$ are found using (1.11) and (1.9) as:

$$\begin{aligned} &\hat{r}_0 = \frac{1}{r_0}, \qquad \hat{r}_1 = \frac{-(r_3 r_4 + r_1 r_5)/r_0}{r_5(r_0 + r_1) + r_3 r_4}, \qquad \hat{r}_2 = \frac{-r_2/r_0}{r_0 + 2/3 r_2} \\ &\hat{r}_3 = \frac{-r_3}{r_5(r_0 + r_1) + r_3 r_4}, \quad \hat{r}_4 = \frac{-r_4}{r_5(r_0 + r_1) + r_3 r_4}, \quad \hat{r}_5 = \frac{r_0 + r_1}{r_5(r_0 + r_1) + r_3 r_4} \end{aligned} \tag{1.12}$$

With the use of inequalities ($1.3_{1,2,3,4}$) and (1.12) it is checked directly that $\mathbf{N}_r^{-1}(\underline{n})$ is a $N^+$-operator. If $\mathbf{N}_r(\underline{n})$ is positive, $\mathbf{N}_r^{-1}(\underline{n})$ is positive too.

(iii) The product $\mathbf{N}_p(\underline{n}) = \mathbf{N}_r(\underline{n})\bullet\mathbf{N}_q(\underline{n})$ of two $N^+$ operators is $N^+$ operator, because $\forall \mathtt{r},\mathtt{q} \in R_6^+ : \; \mathtt{p} \in R_6^+$. This follows from the direct calculations with the use of (1.9):

$$\begin{aligned} &p_0 = q_0 r_0 > 0, \;\; 3/2\, p_0 + p_2 = 3/2(q_0 + 2/3 q_2)(r_0 + 2/3 r_2) > 0 \\ &p_5(p_0 + p_1) + p_4 p_3 = [q_5(q_0 + q_1) + q_4 q_3][r_5(r_0 + r_1) + r_4 r_3] > 0 \end{aligned}. \tag{1.13}$$

The properties of $N^+$-operators established in (i)-(iii) prove theorem 2.

*Remark 2.1*

Due to (1.9), $p_0 + p_1 = (q_0 + q_1)(r_0 + r_1) - q_3 r_4$, $p_5 = q_5 r_5 - q_4 r_3$, therefore the product of two positive $N^+$-operators is *positive* only under additional constraints: $q_3 r_4,\, q_4 r_3 < 0$.

*Remark 2.2*

As a consequence of Remark 2.1, the product of two *positive* ON-operators is generally $N^+$-operator, which is *positive* only under additional constraint, $q_3 r_3 > 0$.

*Remark 2.3*

The multiplicative group $\mathrm{N}_6$ of $N^+$-operators also constitutes the *additive group* relative to the addition of its elements, where $\mathbf{N}_q(\underline{n}) + \mathbf{N}_r(\underline{n}) = \mathbf{N}_{q+r}(\underline{n})$. Therefore the set $N^+$-operators constitutes the *associative ring* $\acute{\mathrm{N}}_6$ relative to both, addition and multiplication operations. In the following we are interested only in the multiplicative properties of N-operators.

*Example*: Dual N-operations

Two linear nematic transformations, $\underline{\underline{z}} = \mathbf{N}_q(\underline{n}) \bullet \underline{\underline{y}} = \mathbf{N}_r(\underline{n}) \bullet \underline{\underline{x}}$ where $\mathbf{N}_r(\underline{n})$ and $\mathbf{N}_q(\underline{n})$ are N-operators, are called *dual.* When both the $\mathbf{N}_r(\underline{n})$ and $\mathbf{N}_q(\underline{n})$ are of $N^+$ type, there always are the unique dependences, $\underline{\underline{y}} = \mathbf{N}_p(\underline{n}) \bullet \underline{\underline{x}}$ and $\underline{\underline{x}} = \mathbf{N}_{\hat{p}}(\underline{n}) \bullet \underline{\underline{y}}$ where

$$\mathbf{N}_p(\underline{n}) = \mathbf{N}_q^{-1}(\underline{n}) \bullet \mathbf{N}_r(\underline{n}), \quad \mathbf{N}_{\hat{p}}(\underline{n}) = \mathbf{N}_p^{-1}(\underline{n}) = \mathbf{N}_r^{-1}(\underline{n}) \bullet \mathbf{N}_q(\underline{n}). \tag{1.14}$$

Formulae (1.9) and (1.12) express the basis scalars $\mathrm{p}$ and $\hat{\mathrm{p}}$ for dual $N^+$-operators via given basis scalars $\mathrm{r}$ and $\mathrm{q}$. In case $\mathbf{N}_p(\underline{n}) = \mathbf{N}_q^{o-1}(\underline{n}) \bullet \mathbf{N}_r^o(\underline{n})$, where $\mathbf{N}_r^o(\underline{n})$ and $\mathbf{N}_q^o(\underline{n})$ are *positive* ON-operators, the parameters $\mathrm{p}$ are:

$$\begin{aligned} &p_0 = \frac{r_0}{q_0}, \qquad p_1 = \frac{(r_0 + r_1)q_5 - r_3 q_3}{q_5(q_0 + q_1) - q_3^2} - \frac{r_0}{q_0}, \qquad p_2 = \frac{r_2 q_0 - r_0 q_2}{q_0(q_0 + 2/3 q_2)}, \\ &p_3 = \frac{r_3(q_0 + q_1) - q_3(r_0 + r_1)}{q_5(q_0 + q_1) - q_3^2}, \quad p_4 = \frac{r_5 q_3 - r_3 q_5}{q_5(q_0 + q_1) - q_3^2}, \quad p_5 = \frac{r_5(q_0 + q_1) - r_3 q_3}{q_5(q_0 + q_1) - q_3^2} \end{aligned} \tag{1.15}$$

Due to (1.14) respective formulae for the basis parameters $\hat{\mathrm{p}}$ of inverse dual $N^+$ operation are obtained from (1.15) by substitution $\mathrm{r} \leftrightarrow \mathrm{q}$. Applying (1.13) to the terms in (1.15) for *positive* ON-operators $\mathbf{N}_r^o(\underline{n})$ and $\mathbf{N}_q^o(\underline{n})$, shows that in this case the *sufficient condition* for $\mathbf{N}_p(\underline{n})$ to be positive is: $r_3 q_3 < 0$.

**1.4. Spectral properties of N-operators**

This Section considers the case of *non-degenerating* (NG) $\mathbf{N}_r(\underline{n})$ operators whose basis parameters $r$ do not vanish:

$$r_k \neq 0 \ \ (k = 0,1,2,3,4,5)\,. \qquad (1.16)$$

The spectral problem for a NG operator $\mathbf{N}_r(\underline{n})$ is formulated in the standard way:

$$\mathbf{N}_r(\underline{n})\bullet \underline{\underline{x}} = \nu \underline{\underline{x}}\,, \ \text{or} \ [\mathbf{N}_r(\underline{n}) - \nu \mathbf{I}(\underline{n})]\bullet \underline{\underline{x}} = \underline{\underline{0}}\,. \qquad (1.17_1)$$

Here $\mathbf{N}_r(\underline{n}) = \sum_{k=0}^{5} r_k \mathbf{a}_k(\underline{n})$, $\mathbf{I}(\underline{n}) = \mathbf{a}_0 + \mathbf{a}_5(\underline{n})$, $\nu$ being a generally complex eigenvalue, and $\underline{\underline{x}}(\nu) \in \breve{X}$ is a respective "eigentensor".

*Theorem 3*

(i) The spectral points (egenvalues) of problem ($1.17_1$) for any NG operator $\mathbf{N}_r(\underline{n})$, are:

$$\nu_1 = r_0, \quad \nu_2 = r_0 + 2/3 r_2, \quad \nu_3 = 1/2(r_0 + r_1 + r_5 + d)\ , \quad \nu_4 = 1/2(r_0 + r_1 + r_5 - d)$$

$$d^2 = (r_0 + r_1 + r_5)^2 - 4[r_5(r_0 + r_1) + r_3 r_4] \equiv (r_0 + r_1 - r_5)^2 - 4 r_3 r_4\,. \qquad (1.18)$$

(ii) The corresponding eigentensors $\underline{\underline{x}}(\nu_k)$ are found as $\underline{\underline{x}}(\nu_k) = \mathbf{Q}(\nu_k, \underline{n})\bullet \underline{\underline{x}}_0$ where $\underline{\underline{x}}_0$ is a given tensor, and the "eigenoperators" $\mathbf{Q}(\nu_k, \underline{n})$ are:

$$\mathbf{Q}(\nu_1, \underline{n}) = c_1(\mathbf{a}_0 - \mathbf{a}_1 - 3/2\mathbf{a}_2), \ \ \mathbf{Q}(\nu_2, \underline{n}) = c_2 \mathbf{a}_2, \ \ \mathbf{Q}(\nu_3, \underline{n}) = c_3(\mathbf{a}_1 - \lambda_1 \mathbf{a}_3) + c_4(\mathbf{a}_4 + \lambda_1 \mathbf{a}_5)$$

$$\mathbf{Q}(\nu_4, \underline{n}) = c_5(\mathbf{a}_1 - \lambda_2 \mathbf{a}_3) + c_6(\mathbf{a}_4 + \lambda_2 \mathbf{a}_5) \ \ \{\lambda_1 = (r_0 + r_1 - \nu_3)/r_4,\ \lambda_2 = (r_0 + r_1 - \nu_4)/r_4\} \quad (1.19)$$

(iii) In case of the *dual* operator $\mathbf{N}_p(\underline{n})$ consisting of two *ON positive* operators, with the parameters $p_k$ defined in (1.15), all eigenvalues in (1.18) are real positive.

*Proof*

(i) A common, analytical continuation $\nu \to \hat{\nu}$ is used to find eigenvalues. Substituting $r_0 \to r_0 - \hat{\nu}$ and $r_5 \to r_5 - \hat{\nu}$, and using (1.12) results in formal finding the basis parameters $\hat{r}_k(\hat{\nu}, r_k)$ of operator $\mathbf{N}_r^{-1}(\hat{\nu}, \underline{n})$. The eigenvalues (1.18) are then found as singular points $\hat{\nu} = \nu$ for the basis parameters $\hat{r}_k(\hat{\nu}, r_k)$ of the inverse operator $\mathbf{N}_r^{-1}(\hat{\nu}, \underline{n})$.

(ii) Instead of $\underline{\underline{x}}(\nu)$, the N-"eigenoperator" $\mathbf{Q}(\nu, \underline{n}) = \sum_{k=0}^{5} q_k(\nu)\mathbf{a}_k$ with known eigenvalues $\nu_k$, is now searched from the equation:

$$\mathbf{N}_r(\underline{n})\bullet\mathbf{Q}(\nu, \underline{n}) - \nu\mathbf{Q}(\nu, \underline{n}) \equiv \mathbf{N}_r(\nu, \underline{n})\bullet\mathbf{Q}(\nu, \underline{n}) = \mathbf{0}\,; \ \mathbf{N}_r(\nu, \underline{n}) = \mathbf{N}_r(\underline{n}) - \nu\mathbf{I}(\underline{n})\,. \qquad (1.17_2)$$

Evidently, the tensor $\underline{\underline{x}} = \mathbf{Q}(\nu,\underline{n})\bullet\underline{\underline{x}}_0$ with $\underline{\underline{x}}_0$ being any given tensor is an eigentensor, because it identically satisfies (1.23$_2$). To find the solution (1.23$_2$) one can use (1.9) with $r_0 \to r_0 - \nu$, $r_5 \to r_5 - \nu$. Demanding then due to (1.17$_2$) $\mathrm{p} = 0$, yields:

$$(r_0+\nu)q_0 = 0,\ \ (r_0+r_1-\nu)q_1 + r_1 q_0 - r_4 q_3 = 0,\ \ (r_0+2/3r_2-\nu)q_2 + r_2 q_0 = 0$$
$$(q_0+q_1)r_3 + q_3(r_5-\nu) = 0,\ \ (r_0+r_1-\nu)q_4 + r_4 q_5 = 0,\ \ -r_3 q_4 + (r_5-\nu)q_5 = 0$$

For each value of $\nu$ from (1.18), the parameters $\mathrm{q}(\nu)$ of eigenoperator $\mathbf{Q}(\nu,\underline{n})$ are found from the above equations as:

$$q(\nu_1) = c_1(1,-1,-3/2,0,0,0),\ q(\nu_2) = c_2(0,0,1,0,0,0),\ q(\nu_3) = (0,c_3,0,\lambda_1 c_3,c_4,-\lambda_1 c_4)$$
$$q(\nu_4) = (0,c_5,0,\lambda_2 c_5,c_6,-\lambda_2 c_6)\quad \{\lambda_1 = (r_0+r_1-\nu_3)/r_5,\ \lambda_2 = (r_0+r_1-\nu_4)/r_5\ \} \qquad (1.20)$$

Here $c_1, c_2, ..., c_6$ are arbitrary constants. Additional solutions that use specific relations between parameters $r_k$ have been rejected because they are *not robust*.

(iii) In case of the *dual* operator $\mathbf{N}_p(\underline{n})$ consistent of two *ON positive* operators, one should use in (24) the substitution $r_k \to p_k(r_i, q_j)$ with $p_k$ presented in (1.15).

The proof is given in the following steps:

1) $\nu_1 = p_0 = r_0/q_0 > 0$; 2) $\nu_2 = p_0 + 2/3p_2 = (r_0 + 2/3r_2)/(q_0 + 2/3q_2) > 0$;

3) $p_0 + p_1 + p_5 = \dfrac{(r_0+r_1)q_5 + (q_0+q_1)r_5 - 2r_3q_3}{q_5(q_0+q_1) - q_3^2} > 0$, because

$$(r_0+r_1)q_5 + (q_0+q_1)r_5 - 2q_3r_3 \ge 2[(r_0+r_1)r_5(q_0+q_1)q_5]^{1/2} - 2q_3r_3 > 2(|q_3r_3| - q_3r_3);$$

4) $d^2 = \dfrac{[(r_0+r_1)q_5 - (q_0+q_1)r_5]^2 + 4\Phi}{[q_5(q_0+q_1) - q_3^2]^2}$, $\quad \Phi = r_5(r_0+r_1)r_3^2 g(x)$,

$$g(x) = (x-\alpha)(x-\beta)\quad (x \equiv q_3/r_3),\ \alpha = (q_0+q_1)/(r_0+r_1),\quad \beta = q_5/r_5 .$$

5) $\min g(x) = g(x_m) = g\{1/2(\alpha+\beta)\} = -1/4(\alpha-\beta)^2$.

6) $d^2 \ge \dfrac{[(r_0+r_1)q_5 - (q_0+q_1)r_5]^2 + 4\min\Phi}{[q_5(q_0+q_1) - q_3^2]^2} = \dfrac{[(r_0+r_1)q_5 - (q_0+q_1)r_5]^2[q_5(q_0+q_1) - q_3^2]}{r_5(r_0+r_1)[q_5(q_0+q_1) - q_3^2]^2} > 0.$

*While the above dual* $\mathbf{N}_p(\underline{n})$ *operator is generally not positive, its eigenvalues are positive.*

*Remark 3.1*

Along with the part (iii) of Theorem 3.1, easy analysis reveals various cases of behavior of eigenvalues $\nu_k$ in (1.18).

(i) In general case of NG operators $\mathbf{N}_r(\underline{n})$, the egenvalues $\nu_1, \nu_2$ are real but generally have arbitrary signs, the egenvalues $\nu_3, \nu_4$ being generally complex and conjugated.

(ii) In case of N$^+$ operators when $\mathbf{N}_r(\underline{n}) \in$ N$_6$, the egenvalues $\nu_1, \nu_2$ are positive, the egenvalues $\nu_3, \nu_4$ being generally complex and conjugated.

(iii) In case of *positive* N$^+$ operators, the egenvalues $\nu_1, \nu_2$ are positive, the egenvalues $\nu_3, \nu_4$ being generally complex and conjugated, with $\mathrm{Re}(\nu_3, \nu_4) > 0$. All eigenvalues $\nu_k$ in (1.24) are real positive if $r_3 r_4 < 0$, as in particular case of positive ON operators where $r_4 = -r_3$.

*Remark 3.2*

Arbitrary parameters $c_k$ in (1.15) can be established from various physical conditions. One of them is:

$$\sum_{i=1}^{4} \mathbf{Q}(\nu_i, \underline{n}) = \mathbf{I}(\underline{n}) = \mathbf{a}_0 + \mathbf{a}_5(\underline{n}). \qquad (1.21)$$

Using (1.20) and (1.21) yields:

$$c_1 = 1, \;\; c_2 = 3/2, \;\; c_3 = (r_0 + r_1 - \nu_4)/d, \;\; c_4 = -r_4/d, \;\; c_5 = (\nu_3 - r_0 - r_1)/d, \;\; c_6 = r_4/d \qquad (1.22)$$

Here parameters $r_k, \nu_3, \nu_4$, and $d$ have been defined in (1.18).

## 1.5. Symmetric N-operators: Transversal isotropy (TI)

### *1.5.1. General properties*

This Section briefly describes the nematic operations on the subgroup $\breve{X}_s \subset \breve{X}$ of traceless second rank symmetric tensors $\underline{\underline{x}}_s \in \breve{X}_s$: $tr\,\underline{\underline{x}}_s = 0$. Therefore the lower index "s" is omitted here. The linear symmetric N-operator $\mathbf{S}_r(\underline{n})$ on $\breve{X}_s$ is defined as:

$$\underline{\underline{y}} = \mathbf{S}_r(\underline{n}) \bullet \underline{\underline{x}} = \sum_{k=0}^{2} r_k \mathbf{a}_k(\underline{n}) \bullet \underline{\underline{x}}, \quad \text{or} \quad \mathbf{S}_r(\underline{n}) = \sum_{k=0}^{2} r_k \mathbf{a}_k(\underline{n}) \qquad (1.23)$$

Here $\mathbf{a}_k(\underline{n})$ are the basis tensors defined in (1.4), and $r_k$ are the real-valued basic scalar ordered parameters $\{r\}$, characterizing operation. The common tensor presentation of symmetric operation (1.1), and corresponding quadratic form $P_s$ are:

$$\underline{\underline{y}} = r_0 \underline{\underline{x}} + r_1[\underline{nn} \cdot \underline{\underline{x}} + \underline{\underline{x}} \cdot \underline{nn} - 2nn(\underline{\underline{x}} : \underline{nn})] + r_2(\underline{nn} - \underline{\underline{\delta}}/3)(\underline{\underline{x}} : \underline{nn}) . \qquad (1.24)$$

$$P_s \equiv \underline{\underline{x}} \bullet \mathbf{S}_r(\underline{n}) \bullet \underline{\underline{x}} \equiv r_0 \left| \underline{\underline{x}}^s \right|^2 + 2r_1 \underline{nn} : \underline{\underline{x}}_s^2 + (r_2 - 2r_1)(\underline{nn} : \underline{\underline{x}}_s)^2 . \qquad (1.25)$$

Equations (1.23) and (1.24) could also be obtained from (1.1) when $y_a \equiv 0$, using the *normalizing procedure* [4,5]. In this case the second equation in (1.1) is used for expressing $x_a$ via $x_s$ and $\underline{n}$. Substituting this dependence $x_a = x_a(x_s, \underline{n})$ into (1) results in the equations (1.23), (1.24). As seen, this procedure does not violate the non-degrading conditions (1.16). Relation (1.30) shows that symmetric N-operators are *transversally isotropic*. Therefore they are called *TI-operators.*

### *1.5.2. Multiplicative group*

Due to the Table1, the products of TI-operators are commutative:

$$\mathbf{S}_p(\underline{n}) = \mathbf{S}_r(\underline{n}) \bullet \mathbf{S}_q(\underline{n}) = \mathbf{S}_q(\underline{n}) \bullet \mathbf{S}_r(\underline{n}) . \qquad (1.26)$$

Here the basis parameters $p_k$ are found from the fundamental equation:

$$p_0 = r_0 q_0, \;\; p_1 = r_0 q_1 + r_1 q_0 + r_1 q_1, \; p_2 = r_0 q_2 + r_2 q_0 + 2/3 r_2 q_2 . \qquad (1.27)$$

A TI-operator is *called positive* if $\forall \underline{\underline{x}} \in \breve{X}$ the quadratic form $P_s = \underline{\underline{x}} \bullet \mathbf{S}_r(\underline{n}) \bullet \underline{\underline{x}} > 0$.

*Theorem 4*

TI operator $\mathbf{S}_r(\underline{n})$ is positive iif

$$r_0 > 0, \;\; r_0 + r_1 > 0, \;\; r_0 + 2/3 r_2 > 0 . \qquad (1.28)$$

*Proof* is the same as for the Theorem 1.

*Theorem 5*

The set of positive TI operators constitutes commutative three parametric TI group $S_3$.

*Proof*

(i) Direct calculations show that the product (1.26) of two positive TI-operators is positive.

(ii) Direct calculations show that the unit TI operation is $\mathbf{I} = \mathbf{a}_0$, so

$$\forall \mathbf{S}_r(\underline{n}): \ \mathbf{I} \bullet \mathbf{S}_r(\underline{n}) = \mathbf{S}_r(\underline{n}) \bullet \mathbf{I} = \mathbf{S}_r(\underline{n}) \ \mathbf{S}_r(\underline{n}).$$

(iii) The basis scalar parameters $\hat{r}_k$ of inverse positive TI-operator $\mathbf{S}_r^{-1}(\underline{n}) \equiv \mathbf{S}_{\hat{r}}(\underline{n})$ are found from the relation $\mathbf{S}_r^{-1}(\underline{n}) \bullet \mathbf{S}_r(\underline{n}) = \mathbf{S}_{\hat{r}}(\underline{n}) \bullet \mathbf{S}_r(\underline{n}) = \mathbf{a}_0(\underline{n})$ as:

$$\hat{r}_0 = \frac{1}{r_0}, \quad \hat{r}_1 = -\frac{r_1 / r_0}{r_0 + r_1}, \quad \hat{r}_2 = -\frac{r_2 / r_0}{r_0 + 2/3 r_2}. \tag{1.29}$$

Due to (1.28), (1.29), any positive TI-operator $\mathbf{S}_r(\underline{n})$ has inverse positive.

The dual linear transformations, $\underline{\underline{z}} = \mathbf{S}_q(\underline{n}) \bullet \underline{\underline{y}} = \mathbf{S}_r(\underline{n}) \bullet \underline{\underline{x}}$ with positive TI operators $\mathbf{S}_q(\underline{n})$ and $\mathbf{S}_r(\underline{n})$ define the direct $\underline{\underline{y}} = \mathbf{S}_p(\underline{n}) \bullet \underline{\underline{x}}$, and inverse, $\underline{\underline{x}} = \mathbf{S}_{\hat{p}}(\underline{n}) \bullet \underline{\underline{y}}$, linear relations, where

$$\mathbf{S}_p(\underline{n}) = \mathbf{S}_q^{-1}(\underline{n}) \bullet \mathbf{S}_r(\underline{n}), \quad \mathbf{S}_{\hat{p}}(\underline{n}) = \mathbf{S}_p^{-1}(\underline{n}) = \mathbf{S}_r^{-1}(\underline{n}) \bullet \mathbf{S}_q(\underline{n}), \tag{1.30}$$

and parameters $\mathrm{p}$ are:

$$p_0 = \frac{r_0}{q_0}, \quad p_1 = \frac{r_1 q_0 - q_1 r_0}{q_0 + q_1}, \quad p_2 = \frac{r_2 q_0 - q_2 r_0}{q_0 + 2/3 q_2}. \tag{1.31$_1$}$$

Parameters $\hat{\mathrm{p}}$ of inverse operation found by substitution $\mathrm{q} \leftrightarrow \mathrm{r}$ are:

$$\hat{p}_0 = \frac{q_0}{r_0}, \quad \hat{p}_1 = \frac{q_1 r_0 - r_1 q_0}{r_0 + r_1}, \quad \hat{p}_2 = \frac{q_2 r_0 - r_2 q_0}{r_0 + 2/3 r_2}. \tag{1.31$_2$}$$

### *1.5.3. Eigenvalue problem*

The formulation of the eigenvalue problem, similar to (1.17$_1$) is:

$$[\mathbf{S}_r(\underline{n}) - \nu \mathbf{I}] \bullet \underline{\underline{x}} = \underline{\underline{0}}, \quad \text{or} \quad \mathbf{S}_r(\underline{n}) \bullet \mathbf{Q}(\nu, \underline{n}) - \nu \mathbf{Q}(\nu, \underline{n}) \equiv \tilde{\mathbf{S}}_r(\nu, \underline{n}) \bullet \mathbf{Q}(\nu, \underline{n}) = \mathbf{0}. \tag{1.32}$$

Here $\mathbf{S}_r(\underline{n}) = \sum_{k=0}^{2} r_k \mathbf{a}_k(\underline{n}) \in \mathrm{S}_3$, $\mathbf{I} = \mathbf{a}_0$, $\mathbf{Q}(\nu, \underline{n}) = \sum_{k=0}^{2} q_k(\nu) \mathbf{a}_k(\underline{n})$ and $\underline{\underline{x}} = \underline{\underline{x}}(\nu) \in \breve{X}_s$.

*Theorem 6*

$\forall \mathbf{N}_r(\underline{n}) \in \mathrm{S}_3$, the spectral points (egenvalues) of problem (1.32) are:

$$\nu_1 = r_0, \quad \nu_2 = r_0 + r_1, \ \nu_3 = r_0 + 2/3 r_2. \tag{1.33}$$

The corresponding eigentensors $\underline{\underline{x}}(\nu_k)$ are found as $\underline{\underline{x}}(\nu_k) = \mathbf{Q}(\nu_k, \underline{n}) \bullet \underline{\underline{x}}_0$ where $\underline{\underline{x}}_0$ is a given tensor, and the "eigenoperators" $\mathbf{Q}(\nu_k, \underline{n})$ are given by:

$$\mathbf{Q}(\nu_1,\underline{n}) = c_1(\mathbf{a}_0 - \mathbf{a}_1 - 3/2\mathbf{a}_2)\,,\quad \mathbf{Q}(\nu_2,\underline{n}) = c_2\mathbf{a}_1\,,\quad \mathbf{Q}(\nu_3,\underline{n}) = c_3\mathbf{a}_2\,. \tag{1.34}$$

*Proof* employs the same technique as used in proof of Theorem 3.

*Remark 6.1*

Due to (1,28), $\nu_k > 0$.

*Remark 6.2*

If the arbitrary parameters $c_k$ in (1.34) are found from the physical condition,

$$\sum_{i=1}^{3}\mathbf{Q}(\nu_i,\underline{n}) = \mathbf{I}(\underline{n}) = \mathbf{a}_0\,, \tag{1.35}$$

their values are:

$$c_1 = c_2 = 1,\quad c_3 = 3/2\,. \tag{1.36}$$

### *1.5.4. Singular TI operators*

Consider now the limiting *marginal situation*, when some inequalities in (1.34) turn out to be equalities. If once again non-degeneration conditions, $r_k \neq 0\ (k = 0,1,2)$, with $r_0 > 0$, are used, there might be only two independent marginal conditions:

$$r_0 + r_1 = 0,\quad r_0 + 2/3r_2 = 0 \tag{1.37$_1$}$$

When one of these conditions is satisfied, TI operator is called *partially soft.* When both of them are satisfied, TI operator is called *completely soft*. In both the partially or complete soft cases, the quadratic form $P_s$ is *positively semi-definite* (*non-negative*).

The *nearly marginal situation* are defined as those that reduce (1.43$_1$) to:

$$r_0 + r_1 = r_0\delta,\quad r_0 + 2/3r_2 = 2/3r_0\kappa\quad (0 < \delta,\kappa << 1);\quad \mathtt{r} = r_0(1,\delta - 1,\kappa - 3/2)\,. \tag{1.37$_2$}$$

Formulae (1.37$_2$) mean that the corresponding TI operator $\mathbf{S}_r(\underline{n}) \equiv \mathbf{S}_{\delta,\kappa}(\underline{n})$ is positive. Due to (1.37$_1$) there is one-parametric *marginal family* of TI operators, $\mathbf{S}(\underline{n}) = r_0\boldsymbol{\alpha}(\underline{n})$ $(r_0 > 0)$ with the parameters $\mathtt{r}$, represented as:

$$\boldsymbol{\alpha}(\underline{n}) = \mathbf{a}_0(\underline{n}) - \mathbf{a}_1(\underline{n}) - 3/2\mathbf{a}_2(\underline{n})\,,\quad \mathtt{r} = r_0(1,-1,-3/2)\ . \tag{1.38}$$

It is seen that the marginal TI operator $\boldsymbol{\alpha}(\underline{n})$ is singular, i.e. $\boldsymbol{\alpha}^{-1}(\underline{n})$ does not exist. The operator $\boldsymbol{\alpha}(\underline{n})$ has the property:

$$\boldsymbol{\alpha}(\underline{n})\bullet\boldsymbol{\alpha}(\underline{n}) = \boldsymbol{\alpha}(\underline{n}). \tag{1.39}$$

Consider now a pair $\mathbf{S}_r(\underline{n}) = \mathbf{S}_{\delta_1,\kappa_1}(\underline{n})$ and $\mathbf{S}_q(\underline{n}) = \mathbf{S}_{\delta_2,\kappa_2}(\underline{n})$ of positive, nearly marginal TI operators, with parameters: $r = (1, \delta_1 - 1, \kappa_1 - 3/2)$ and $q = (1, \delta_2 - 1, \kappa_2 - 3/2)$, where $0 < \delta_1, \kappa_1 << 1$, and $0 < \delta_2, \kappa_2 << 1$. Evidently $\mathbf{S}_{\delta_1,\kappa_1}(\underline{n}) \to \mathbf{S}_{\delta_2,\kappa_2}(\underline{n}) \to \boldsymbol{\alpha}(\underline{n})$ when $(\delta_1, \kappa_1) \to (\delta_2, \kappa_2) \to (0,0)$.

*Theorem $7_1$*

When the two independent limits exist:

$$\delta = \lim_{\delta_{1,2}\to 0} (\delta_2 / \delta_1), \quad \kappa = \lim_{\nu_{1,2}\to 0} (\kappa_2 / \kappa_1), \quad (0 < \delta, \kappa < \infty) \tag{1.40}$$

there exist two positive limiting TI operators,

$$\boldsymbol{\alpha}_{\delta,\kappa}(\underline{n}) \equiv \lim_{\delta_{1,2},\kappa_{1,2}\to 0} \mathbf{S}^{-1}{}_{\delta_1,\kappa_1}(\underline{n}) \bullet \mathbf{S}_{\delta_2,\kappa_2}(\underline{n}) = \mathbf{a}_0(\underline{n}) + (\delta - 1)\mathbf{a}_1(\underline{n}) + 3/2(\kappa - 1)\mathbf{a}_2(\underline{n}) \tag{1.41$_1$}$$

$$\boldsymbol{\beta}_{\delta,\kappa}(\underline{n}) \equiv \lim_{\delta_{1,2},\kappa_{1,2}\to 0} \mathbf{S}^{-1}_{\delta_2,\kappa_2}(\underline{n}) \bullet \mathbf{S}_{\delta_1,\kappa_1}(\underline{n}) = \mathbf{a}_0(\underline{n}) + (\delta^{-1} - 1)\mathbf{a}_1(\underline{n}) + 3/2(\kappa^{-1} - 1)\mathbf{a}_2(\underline{n}). \tag{1.41$_2$}$$

The TI operators defined in ($1.47_{1,2}$) have the following properties:

$$\boldsymbol{\beta}_{\delta,\kappa}(\underline{n}) = \boldsymbol{\alpha}^{-1}_{\delta,\kappa}(\underline{n}) = \boldsymbol{\alpha}_{1/\delta,1/\kappa}(\underline{n}), \tag{1.42$_1$}$$

$$\boldsymbol{\alpha}_{\delta,\kappa}(\underline{n}) \bullet \boldsymbol{\alpha}(\underline{n}) = \boldsymbol{\beta}_{\delta,\kappa}(\underline{n}) \bullet \boldsymbol{\alpha}(\underline{n}) = \boldsymbol{\alpha}(\underline{n}), \tag{1.42$_2$}$$

$$\boldsymbol{\alpha}_{1,1}(\underline{n}) = \boldsymbol{\beta}_{1,1}(\underline{n}) = \mathbf{a_0}(\underline{n}). \tag{1.42$_3$}$$

The *proof* follows from direct calculations using ($1.41_{1,2}$).

*Theorem $7_2$*

When either $\delta \to 0$ and/or $\kappa \to 0$, or $\delta \to \infty$ or/and $\kappa \to \infty$, there are only two non-trivial alternatives of singular limiting behavior of TI operators $\boldsymbol{\alpha}_{\delta,\kappa}(\underline{n})$ and $\boldsymbol{\beta}_{\delta,\kappa}(\underline{n})$:

(i) $\alpha$*-type* described by singular TI operators $\boldsymbol{\alpha}_{0,\kappa}(\underline{n})$ or $\boldsymbol{\alpha}_{\delta,0}(\underline{n})$, or $\boldsymbol{\alpha}_{0,0}(\underline{n}) = \boldsymbol{\alpha}(\underline{n})$, when the operators $\boldsymbol{\beta}_{\delta,\kappa}(\underline{n})$ do not exist; and

(ii) $\beta$*-type* described by singular TI operators $\boldsymbol{\beta}_{\infty,\kappa}(\underline{n})$ or $\boldsymbol{\beta}_{\delta,\infty}(\underline{n})$, or $\boldsymbol{\beta}_{\infty,\infty}(\underline{n}) = \boldsymbol{\alpha}(\underline{n})$, when the operators $\boldsymbol{\alpha}_{\delta,\kappa}(\underline{n})$ do not exist.

The *proof* is based on ($1.42_1$).

The cases $\boldsymbol{\alpha}_{0,0}(\underline{n}) = \boldsymbol{\alpha}(\underline{n})$ and $\boldsymbol{\beta}_{\infty,\infty}(\underline{n}) = \boldsymbol{\alpha}(\underline{n})$ describe, respectively, the limiting *super-soft* behaviors of either $\alpha$- or $\beta$-types.

***The eigenvalue problem for the singular operator*** $\mathbf{S}(\underline{n}) = r_0 \boldsymbol{\alpha}(\underline{n})$ where $\boldsymbol{\alpha}(\underline{n})$ is presented in (1.38), has an easy solution. Due to (1.33) the eigenvalues are:

$$\nu_1 = r_0, \quad \nu_1 = \nu_2 = 0, \tag{1.43}$$

whereas the the "eigenoperators" $\mathbf{Q}(\nu_k, \underline{n})$ are given by (1.34). It is also possible to use further the formulae (1.35) and (1.36).